\documentclass[twocolumn,preprintnumbers,amsmath,amssymbm,prd]{revtex4}
\usepackage{epsfig}
\usepackage{graphicx}

\begin{document}
\title{Hyperentropic systems and the generalized second law of thermodynamics}
\author{Shahar Hod}
\affiliation{The Ruppin Academic Center, Emeq Hefer 40250, Israel}
\affiliation{ } \affiliation{The Hadassah Institute, Jerusalem
91010, Israel}
\date{\today}

\begin{abstract}
\ \ \ The holographic bound asserts that the entropy $S$ of a system
is bounded from above by a quarter of the area ${\cal A}$ of a
circumscribing surface measured in Planck areas: $S\leq {\cal
A}/4{\ell^2_P}$. This bound is widely regarded a desideratum of any
fundamental theory. Moreover, it was argued that the holographic
bound is necessary for the validity of the generalized second law
(GSL) of thermodynamics. However, in this work we explicitly show
that hyperentropic systems (those violating the holographic entropy
bound) do exist in higher-dimensional spacetimes. We resolve this
apparent violation of the GSL and derive an upper bound on the area
of hyperentropic objects.
\end{abstract}
\bigskip
\maketitle


The past four decades have witnessed a breakthrough in computer and
data storage technology \cite{BekSch1}. The impressive reduction in
size of information storage devices is one of the most remarkable
advances in this field. It is believed that individual atoms and
molecules may one day become short term information-storage devices
\cite{BekSch1}. Can this trend of miniaturization continue
indefinitely? One (see \cite{BekSch1,Bek1}) naturally wonders if
there is some fundamental limitation on the size devices of given
information capacity may reach in the future? As emphasized by
Bekenstein \cite{BekSch1,Bek1}, such question is related to the
following question: what are the limitations on the magnitude of the
entropy of a system characterized by general parameters such as size
and energy \cite{Bek1}?

The celebrated holographic principle of 't Hooft \cite{Hoof,Suss}
asserts that there is a deep connection between the physical content
of a theory defined in a spacetime and the corresponding content of
another theory defined on the boundary of the same spacetime
\cite{Bek2}. A consistency requirement on the holographic principle
is that the boundary of any physical system should be able to encode
as much information as required to enumerate all possible quantum
states of the bulk system \cite{Bek2}. In light of the
correspondence between information and entropy \cite{inf1}, and the
well-known entropy-area relation for black holes \cite{Bek5,Haw1},
this requirement has been translated into the holographic entropy
bound \cite{Hoof,Suss,Bek2}. This bound asserts that the entropy $S$
(or information) that can be contained in a physical system is
bounded in terms of the area ${\cal A}$ of a surface enclosing it
\cite{Hoof,Suss,Bek2}:
\begin{equation}\label{Eq1}
S\leq {{\cal A}\over{4\ell^2_P}}\  ,
\end{equation}
where $\ell^2_P={G\hbar/c^3}$ is the Planck area. [We shall
henceforth use natural units in which $G=c=k_B=1$.]

The holographic principle \cite{Hoof,Suss} and the holographic bound
(\ref{Eq1}) are widely regarded as guidelines to the ultimate
physical theory of nature \cite{Bek3}. For systems in three spatial
dimensions, the bound (\ref{Eq1}) suggests an information content
that scales no faster than the area of the boundary of the space.
The holographic bound also implies that $(3+1)$-dimensional black
holes have the largest possible entropy among (stationary and
bounded) physical systems characterized by a given surface area
${\cal A}$ (see also \cite{Bek4,Hodspin,BekMayHod}).

As support for the holographic bound in {\it three} spatial
dimensions, Susskind \cite{Suss,Bek2,Wald} described the following
gedanken experiment: Take a neutral nonrotating spherical object of
radius $R$, energy $E$ (with $R>2E$), and entropy $S$ which violates
the holographic bound: $S>\pi R^2/\hbar$. A spherically symmetric
and concentric shell of mass $R/2-E$ is dropped on the system;
according to Birkhoff's theorem the total mass is now $R/2$. When
the outermost surface of the shell reaches Schwarzschild radial
coordinate $r=R$, the system becomes a black hole of radius $R$ and
entropy $S_{\text{BH}}=\pi R^2/\hbar$, which is {\it smaller} than
the original entropy $S$ \cite{Bek2}. Susskind argued that the
apparent violation of the generalized second law of thermodynamics
(GSL) \cite{Bek5} in this gedanken experiment should be regarded as
evidence that the envisaged system cannot really exist.

Before we proceed, it should be mentioned that it is possible to
find examples for systems which violate the holographic bound. For
example, a collapsed object already inside its own gravitational
radius eventually violates it. The enclosing area can only decrease
while the enclosed entropy can only grow \cite{Bek2,Bous1,Yurt}.
Another example is given by a large spherical section of a flat
Friedmann universe: its enclosing area grows like radius squared
while the enclosed entropy does so like radius cubed. These examples
belong to a class of strongly self-gravitating and dynamical
systems. The second example also describes an unisolated system.
Nevertheless, it has been established \cite{Bek2} that the
holographic bound (\ref{Eq1}) can be trusted for generic weakly
self-gravitating isolated systems in three spatial dimensions. In
the present work we shall focus on such weakly self-gravitating
isolated systems.

Clearly, Susskind's gedanken experiment can also be applied to
physical systems in higher-dimensional spacetimes. The arguments of
\cite{Suss} thus suggest that the holographic bound (\ref{Eq1}) must
follow from the GSL in any number of spatial dimensions. But is the
holographic bound really valid for physical systems in {\it
higher}-dimensional spacetimes?

Proliferation of large spatial dimensions is expected to increase
the entropy content of a physical system which is characterized by a
given amount of energy. Evidently the more the dimensions, the more
ways there are to split up a given amount of energy between the
quantum states of the system \cite{Bek6}. Thus, one may expect the
challenge to the holographic entropy bound to become more and more
serious as the number of spatial dimensions increases. As an
example, consider in $D$ flat spatial dimensions a spherical box of
radius $R$ which contains massless fields. We shall follow the
analysis of \cite{Bek6} in order to calculate the system's entropy
in the thermodynamic regime.

The mean thermal energy in the sphere from one helicity degree of
freedom is \cite{Bek6}
\begin{equation}\label{Eq2}
E_{\text{d.o.f}}=V_D(R){\int_0^{\infty}}{{\hbar\omega\
dV_D(\omega)}\over{{(e^{\beta\hbar\omega}\mp 1)}{(2\pi)^D}}}\  ,
\end{equation}
where the upper (lower) signs correspond to boson (fermion) fields,
and $\beta\equiv 1/T$ is the inverse temperature of the system. Here
\begin{equation}\label{Eq3}
V_D(R)={{2\pi^{D/2}}\over{D\Gamma(D/2)}}R^D\
\end{equation}
is the volume of a sphere of radius $R$ in $D$ spatial dimensions,
and
\begin{equation}\label{Eq4}
dV_D(\omega)=[2\pi^{D/2}/\Gamma(D/2)]\omega^{D-1}d\omega\
\end{equation}
is the volume in frequency space of the shell
$(\omega,\omega+d\omega)$. We note that the distribution
$\omega^{D}/(e^{\beta\hbar\omega}\mp 1)$ in Eq. (\ref{Eq2}) peaks at
the characteristic frequency
\begin{equation}\label{Eq5}
\bar\omega={{D}\over{\hbar\beta}}[1\mp e^{-D}+O(e^{-2D})]\  .
\end{equation}

From Eqs. (\ref{Eq2})-(\ref{Eq4}) and the relation
\begin{equation}\label{Eq6}
{\int_0^{\infty}}{{x^Ddx}\over{e^x\mp1}}=\zeta(D+1)\Gamma(D+1)\times
\begin{cases}
1 & \text{for bosons} \ ; \\
1-2^{-D} & \text{for fermions} \ ,
\end{cases}
\end{equation}
where $\zeta(z)$ is the Riemann zeta function, one finds that the
mean energy of all massless fields is given by
\begin{equation}\label{Eq7}
E={{2N\zeta(D+1)\Gamma({{D+1}\over{2}})R^D}\over{{\pi}^{1/2}\Gamma({D\over
2})\beta^{D+1}\hbar^D}}\ ,
\end{equation}
where $N$ is the number of massless degrees of freedom (the number
of polarization states). Massless scalars contribute $1$ to $N$,
massless fermions contribute $1-2^{-D}$ to $N$ \cite{Bek6}, an
electromagnetic field contributes $D-1$ to $N$ \cite{CarCavHal}, and
the graviton contributes $(D+1)(D-2)/2$ to $N$ \cite{CarCavHal}.
Solving Eq. (\ref{Eq7}) for $\beta\hbar/R$ one finds
\begin{equation}\label{Eq8}
\beta\hbar/R=C_D(N\hbar/RE)^{{1}\over{D+1}}\  ,
\end{equation}
where
\begin{equation}\label{Eq9}
C_D\equiv\Big[{{2\zeta(D+1)\Gamma({{D+1}\over{2}})}\over{\pi^{1/2}\Gamma({D\over
2})}}\Big]^{{1}\over{D+1}}\ .
\end{equation}

Likewise, one can write the thermal entropy of one helicity degree
of freedom as \cite{Bek6}
\begin{equation}\label{Eq10}
S_{\text{d.o.f}}=V_D(R){\int_0^{\infty}} \Big[\mp \ln (1\mp
e^{-\beta\hbar\omega})+{{\beta\hbar\omega}\over{e^{\beta\hbar\omega}\mp
1}}\Big]{{dV_D(\omega)}\over{{(2\pi)}^D}}\ .
\end{equation}
After some algebra we obtain
\begin{equation}\label{Eq11}
S={{2N(D+1)\zeta(D+1)\Gamma({{D+1}\over{2}})R^D}\over{{\pi}^{1/2}D\Gamma({D\over
2})\beta^{D}\hbar^D}}\
\end{equation}
for the total entropy of the system. Comparing (\ref{Eq7}) and
(\ref{Eq11}), one deduces the relation
\begin{equation}\label{Eq12}
S={{D+1}\over{D}}\beta E\  .
\end{equation}
Substituting Eq. (\ref{Eq8}) into Eq. (\ref{Eq12}), one finds
\begin{equation}\label{Eq13}
S=C_D(1+1/D)N^{1\over{D+1}}{(RE/\hbar)}^{D\over{D+1}}\
\end{equation}
for the $(D+1)$-dimensional radiation entropy.

It is important to emphasize that our analysis is appropriate only
for weakly self-gravitating systems. In particular, formula
(\ref{Eq13}) for the entropy can be trusted provided the system's
energy (for a given radius $R$) is bounded from above as here
stated. The spacetime outside the spherical box (for $D\ge 3$) is
described by the $(D+1)$-dimensional Schwarzschild-Tangherlini
metric \cite{SchTang,Kun} of ADM energy $E$:
\begin{equation}\label{Eq14}
ds^2=-H(r)dt^2+{H(r)}^{-1}dr^2+r^2d\Omega^{(D-1)}\ ,
\end{equation}
with
\begin{equation}\label{Eq15}
H(r)=1-{\Big({r_g\over r}\Big)}^{D-2}\  .
\end{equation}
Here
\begin{equation}\label{Eq16}
r_g={\Big[{{16\pi E}\over{(D-1)A_{D-1}}}\Big]}^{1\over{D-2}}
\end{equation}
is the gravitational radius of the system and
\begin{equation}\label{Eq17}
A_{D-1}={{2\pi^{D/2}}\over {\Gamma(D/2)}}
\end{equation}
is the area of a unit $(D-1)$-sphere.

For the system to be weakly self-gravitating, one should impose the
criterion $H(r=R)\simeq 1$ at the surface of the sphere, or
equivalently $(r_g/R)^{D-2}\ll 1$. Taking cognizance of Eqs.
(\ref{Eq15})-(\ref{Eq16}), this condition yields the restriction
\begin{equation}\label{Eq18}
RE\ll {{D-1}\over{16\pi}}{\cal A}\  ,
\end{equation}
where ${\cal A}=A_{D-1}R^{D-1}$ is the surface area of the system.
We characterize this restriction by the dimensionless control
parameter $\eta$ defined by
\begin{equation}\label{Eq19}
\eta\equiv {{16\pi RE}\over{(D-1){\cal A}}}\ll 1\  .
\end{equation}

Taking cognizance of Eqs. (\ref{Eq13}) and ({\ref{Eq19}), we can
write the system's thermal entropy as
\begin{equation}\label{Eq20}
S=C_D(1+1/D)N^{1\over{D+1}}{\Big[{{\eta(D-1){\cal
A}}\over{16\pi\hbar}}\Big]}^{D\over{D+1}}\ .
\end{equation}
From Eq. (\ref{Eq20}) one learns that for the system's entropy to
beat the holographic bound (that is, $S>{\cal A}/4\hbar$), its
surface area must be bounded from above according to
\begin{equation}\label{Eq21}
{{\cal
A}\over\hbar}<{[4C_D(1+1/D)]}^{D+1}N{\Big[{{\eta(D-1)}\over{16\pi}}\Big]}^{D}\
.
\end{equation}

The validity of the thermodynamic description rests on the
assumption that {\it many} quanta (of each degree of freedom) are
thermally excited in the system: $E/N\gg \hbar\bar\omega$. Taking
cognizance of Eqs. (\ref{Eq5}) and (\ref{Eq7}), one finds the
thermodynamic condition
\begin{equation}\label{Eq22}
{C_D}^{D+1}{({{R}/{\beta\hbar}})}^D\gg D\  .
\end{equation}
Using the relation (\ref{Eq8}), one can cast this condition in the
form
\begin{equation}\label{Eq23}
{C^{-1}_D}{(N\hbar/RE)}^{D\over{D+1}}\ll D^{-1}\  .
\end{equation}
We characterize this constraint by the dimensionless control
parameter $\xi$ defined by
\begin{equation}\label{Eq24}
\xi\equiv C^{-1}_D D{(N\hbar/RE)}^{D\over{D+1}}\ll 1\  .
\end{equation}

Solving Eqs. (\ref{Eq19}) and (\ref{Eq24}) for $RE$, one can express
the system's area as
\begin{equation}\label{Eq25}
{{\cal A}\over{\hbar}}={{16\pi N D^{{D+1}\over
D}}\over{\eta(D-1)(\xi C_D)^{{D+1}\over D}}}\  .
\end{equation}
Substituting (\ref{Eq25}) into (\ref{Eq21}), we realize that a
violation of the holographic bound can occur if the number of
spatial dimensions satisfies the inequality \cite{Notecd}
\begin{equation}\label{Eq26}
D\ge D^*\simeq 4\pi/\eta\xi^{1/D}\  .
\end{equation}
Since the dimensionless control parameters satisfy the relations
$\xi\ll 1$ and $\eta\ll 1$ [see Eqs. (\ref{Eq19}) and (\ref{Eq24})],
we learn from (\ref{Eq26}) that the critical dimension $D^*$ (the
minimal value of $D$ above which a violation of the holographic
bound can be realized) satisfies $D^*\gg4\pi$ \cite{Notethree}.

Note that in the large $D$ regime (\ref{Eq26}), the condition
(\ref{Eq21}) for a violation of the holographic bound (\ref{Eq1})
can be simplified:
\begin{equation}\label{Eq27}
{{\cal
A}\over\hbar}<32N\eta^D\sqrt{{\pi}\over{2D}}\Big({{D}\over{4\pi}}\Big)^{D+1}\
.
\end{equation}

Our analysis thus reveals that the holographic bound (\ref{Eq1}) can
actually  be violated in higher-dimensional spacetimes. But
according to the arguments of Ref. \cite{Suss}, the holographic
bound is necessary for the validity of the GSL, for otherwise the
resulting black hole (after collapsing a shell upon the
hyperentropic object) would have an entropy $S_{\text{BH}}={\cal
A}/4\hbar$ which is smaller than the entropy of the original
(hyperentropic) system. We must therefore ask what was wrong with
the original arguments of Susskind \cite{Suss} suggesting that the
holographic bound should follow from the GSL in any number of
spatial dimensions. Can we resolve this apparent violation of the
GSL?

We argue that one can escape a violation of the GSL {\it if} it
turns out that it is actually not possible to form a stable (or
meta-stable) black hole in the gedanken experiment of
\cite{Suss,Wald}. Due to Hawking evaporation \cite{Haw1}, a purely
quantum effect, the black hole will have a {\it finite} lifetime. If
it turns out that this lifetime is shorter than the relaxation time
of the dynamically formed black hole, than a static (or
quasi-static) black hole will never actually form in the gedanken
experiment of \cite{Suss,Wald}. Instead, the intermediate
non-equilibrium configuration will merely act as a catalyst for
transforming the initial high entropy confined state into a final
higher entropy state of unconfined Hawking radiation \cite{Wald}. We
shall now provide analytical estimates for the lifetime,
$\tau_{\text{bh}}$, and dynamical relaxation time,
$\tau_{\text{relaxation}}$, of $(D+1)$-dimensional black holes.

The Hawking radiation power emitted by a $(D+1)$-dimensional black
hole of radius $r_H$ can be approximated by the blackbody formula
\cite{Myers,Kanti1,Hodbul}
\begin{equation}\label{Eq28}
P_D=\sigma_D A_{\text{abs}} T^{D+1}\  ,
\end{equation}
where the $D$-dimensional Stefan-Boltzman constant is given by
\cite{Myers}
\begin{equation}\label{Eq29}
\sigma_D={{A_{D-2}\Gamma(D+1)\zeta(D+1)N}\over{(2\pi)^D(D-1)\Gamma({{D-1}\over{2}})}}\
.
\end{equation}
Here
\begin{equation}\label{Eq30}
A_{\text{abs}}={{A_{D-2}}\over{D-1}}r_c^{D-1}\
\end{equation}
is the absorptive area of the black hole in the geometrical optics
(high energy) limit \cite{Myers,Kanti1}, where
\begin{equation}\label{Eq31}
r_c\equiv
\Big({{D}\over{2}}\Big)^{{1}\over{D-2}}\sqrt{{{D}\over{D-2}}}r_H\ ,
\end{equation}
is the critical radius for null geodesics \cite{Myers,Kanti1} (if a
photon travels inside this radius, it is captured by the black
hole). It was recently shown \cite{Hodbul} that the blackbody
formula (\ref{Eq28}) provides a reasonably good description of the
black-hole emission power. In fact, the agreement between the
(numerically computed) black-hole power and the blackbody analytical
formula (\ref{Eq28}) is very good in the large $D$ regime
(\ref{Eq26}) \cite{Hodbul}.

Substituting Eqs. (\ref{Eq29})-(\ref{Eq31}) into Eq. (\ref{Eq28})
and using the relation \cite{Kun}
$T_{\text{BH}}={{(D-2)\hbar}\over{4\pi r_H}}$ for the black hole's
temperature, one finds
\begin{equation}\label{Eq32}
P_D\simeq
N\Big({{D-2}\over{4\pi}}\Big)^{D+1}\Big({{r_c}\over{r_H}}\Big)^{D-1}{{D\zeta(D+1)\hbar}\over{\pi
r_H^2}}\
\end{equation}
for the total power radiated by a $(D+1)$-dimensional black hole.

The corresponding decrease of the black hole mass during the Hawking
evaporation is given by $d{\cal M}/dt=-P_D$. Using the mass-radius
relation (\ref{Eq16}) in Eq. (\ref{Eq32}), one may integrate this
equation to find
\begin{equation}\label{Eq33}
\tau_{\text{bh}}\simeq
{{2^{2D-2}\pi^{D+1}(D-1)}\over{D^2(D-2)^D\zeta(D+1)\hbar
N}}\Big({{r_H}\over{r_c}}\Big)^{D-1}{\cal A}r_H\
\end{equation}
for the lifetime of the $(D+1)$-dimensional black hole. Note that in
the large $D$ regime (\ref{Eq26}), one has $(r_H/r_c)^{D-1}\simeq
2/D\text{e}$, which implies the compact expression
\begin{equation}\label{Eq34}
\tau_{\text{bh}}\simeq \Big({{4\pi}\over{D}}\Big)^{D+2}{{{\text
e}}\over{32\pi\hbar N}}{\cal A}r_H\  .
\end{equation}

On the other hand, the characteristic timescale required for the
dynamically formed black hole to settle down to a stationary,
equilibrium configuration is given by \cite{Konoq,Hodq,CarBer}
\begin{equation}\label{Eq35}
\tau_{\text{relaxation}}=\Im\omega_0^{-1}={{2\sqrt{D}}\over{D-2}}\Big({{D}\over{2}}\Big)^{{1}\over{D-2}}r_H\
,
\end{equation}
where $\omega_0$ is the fundamental black-hole quasinormal frequency
\cite{Noll}. Note that in the large $D$ regime (\ref{Eq26}), one can
approximate (\ref{Eq35}) by the compact formula
\begin{equation}\label{Eq36}
\tau_{\text{relaxation}}={{2r_H}\over{\sqrt{D}}}\  .
\end{equation}

In order for a $(D+1)$-dimensional black hole to be regarded as a
stable (or meta-stable) state, its lifetime must be longer than its
dynamical relaxation time:
$\tau_{\text{bh}}>\tau_{\text{relaxation}}$. Thus, taking cognizance
of Eqs. (\ref{Eq34}) and (\ref{Eq36}), one may deduce a lower bound
on the area of stable (or meta-stable) $(D+1)$-dimensional
Schwarzschild black holes:
\begin{equation}\label{Eq37}
\Big({{\cal A}\over{\hbar}}\Big)_{\text{min}}\simeq{{16\sqrt{D}
N}\over{\text{e}}}\Big({{D}\over{4\pi}}\Big)^{D+1}\  .
\end{equation}

As discussed above, one can accept a violation of the holographic
entropy bound (the existence of hyperentropic physical systems) and
at the same time avoid a disturbing violation of the GSL in the
gedanken experiment of \cite{Suss,Wald}, provided the lifetime of
the black hole which is formed from the collapse of the
hyperentropic system is shorter than its relaxation time. In this
case, a quasi-static black hole will never actually form in the
gedanken experiment of \cite{Suss,Wald}. Instead, there would be an
intermediate non-equilibrium configuration which will merely act as
a catalyst for converting the initial high entropy confined state
into a final higher entropy state of unconfined Hawking radiation
\cite{Wald}. Taking cognizance of Eq. (\ref{Eq27}), one realizes
that the area of the black hole which would form from the collapse
of the hyperentropic system is {\it smaller} then the minimal area
(\ref{Eq37}) which is required for a meta-stable black-hole
configuration. [The RHS of (\ref{Eq37}) is larger than the RHS of
(\ref{Eq27}) by the factor $\sim D/\eta^D$.] Thus, we conclude that
the GSL is {\it respected} despite the fact that the holographic
bound (\ref{Eq1}) {\it can} be violated.

In summary, the gedanken experiment of \cite{Suss} suggests that the
holographic entropy bound (\ref{Eq1}) is necessary for the validity
of the generalized second law of thermodynamics. However, in this
work we have demonstrated explicitly that the bound (\ref{Eq1}) can
be violated in higher-dimensional spacetimes. At first sight, this
finding seems to open a possibility of violating the GSL in the
gedanken experiment of \cite{Suss}. However, our analysis reveals
that hyperentropic systems are actually allowed to exist (they are
harmless to the GSL) {\it provided} their area is bounded from above
by:
\begin{equation}\label{Eq38}
{{\cal A}\over{\hbar}}<{{16\sqrt{D}
N}\over{\text{e}}}\Big({{D}\over{4\pi}}\Big)^{D+1}\  .
\end{equation}

It is of interest to search for other examples of physical systems
which violate the holographic entropy bound (\ref{Eq1}). It would be
highly important to verify that these systems do conform to the new
area bound (\ref{Eq38}), which is necessary for the validity of the
GSL.

\noindent
{\bf ACKNOWLEDGMENTS}

This research is supported by the Meltzer Science Foundation. I
thank Yael Oren and Arbel M. Ongo for helpful discussions. I thank
Jacob D. Bekenstein for helpful correspondence.



\begin{thebibliography}{99}
\bibitem{Notethree} Note that Eq. (\ref{Eq26}) implies that thermal radiation in three
spatial dimensions conforms to the holographic bound.
\end{thebibliography}
\end{document}